\documentstyle[12pt,aaspp4]{article}

\begin{document}

\def\plottwo#1#2{\centering \leavevmode
\epsfxsize=.5\textwidth \epsfbox[100 150 750 650]{#1} \hfil
\epsfxsize=.5\textwidth \epsfbox[100 150 750 650]{#2}}

\def\etal{{\em et~al.}}
\def\eg{{\em e.g.}}
\def\ie{{\em i.e.}}
\def\vs{{\em vs.}}
\def\degrees{{$^\circ$}}
\def\degree{{$^\circ$}}
\def\mic{$\mu$m}
\def\micron{$\mu$m}
\def\microns{$\mu$m}
\def\kms{km s$^{-1}$}
\def\H2{H$_2$}
\def\lsun{L$_{\odot}$}
\newcommand{\SL}{L_{\odot}}
\def\msun{M$_{\odot}$}
\newcommand{\Lsun}{L_{\odot}}
\newcommand{\Msun}{M_{\odot}}
\newcommand{\SM}{M_{\odot}}
\def\Av{A$_{\mathrm v}$}
\newcommand{\pten}[1]{\times10^{#1}}
\newcommand{\Lstar}{L_{\star}}
\newcommand{\gsim}{\stackrel{>}{\sim}}
\newcommand{\lsim}{\stackrel{<}{\sim}}
\newcommand{\ASCA}{{\it ASCA}}
\newcommand{\ROSAT}{{\it ROSAT}}

\title{X-ray Spectroscopy of the \\ Nearby, Classical T Tauri Star TW Hya}

\author{
        Joel H. Kastner\footnote{Address as of 7/15/99: Rochester Institute of
        Technology, Chester F. Carlson Center for Imaging Science, 84 Lomb
        Memorial Drive, Rochester, NY 14623}, David P. Huenemoerder,
        Norbert S. Schulz \\ 
        \small Center for Space Research, Massachusetts Institute of
Technology, NE80--6007, Cambridge, MA 02139\\ 
        \small e-mail: jhk@juggler.mit.edu
 \and
        \normalsize David A. Weintraub \\
        \small Department of Physics \& Astronomy,
Vanderbilt University, P.O. Box 1807 Station B, Nashville, TN 37235}

\begin{center}
{\em To appear in the {\rm The Astrophysical Journal} \\
(Received 1999 April 28; accepted 1999 May 24)}
\end{center}

\received{1999 April 28}
\accepted{1999 May 24}

\begin{abstract}
We present \ASCA\ and \ROSAT\ X-ray observations of the classical T Tauri
star TW Hya, the namesake of a small association that, at a distance of
$\sim50$ pc, represents the nearest known region of recent star
formation. Analysis of \ASCA\ and \ROSAT\ spectra indicates characteristic
temperatures of $\sim1.7$ MK and $\sim9.7$ MK for the X-ray emitting
region(s) of TW Hya, with emission lines of highly ionized Fe dominating the
spectrum at energies $\sim1$ keV. The X-ray data show variations in X-ray
flux on $\lsim1$ hr timescales as well as indications of changes in X-ray
absorbing column on timescales of several years, suggesting that flares and
variable obscuration are responsible for the large amplitude optical
variability of TW Hya on short and long timescales, respectively.
Comparison with model calculations suggests that TW Hya produces sufficient
hard X-ray flux to produce significant ionization of molecular gas within
its circumstellar disk; such X-ray ionization may regulate both
protoplanetary accretion and protoplanetary chemistry.
\end{abstract}

\section{Introduction}

A primary result of X-ray surveys of star formation regions by the {\it
Einstein} and {\it ROSAT} X-ray satellite observatories is the association
of strong X-ray emission with solar-mass, pre-main sequence (PMS) stars at
various stages of evolution (e.g., Montmerle et al.\ 1993; Feigelson 1996;
and references therein). This result, when viewed in the context of recent
theoretical predictions concerning the potentially profound effects of
X-rays from young stars on the physics and chemistry of their circumstellar,
protoplanetary disks (e.g., Glassgold, Najita, \& Igea 1997; Igea \&
Glassgold 1999), makes clear that knowledge of the detailed X-ray spectra of
pre-main sequence stars is central to an understanding of the early
evolution of the Sun and solar system.  Low-mass PMS stars in Taurus-Auriga,
Chamaeleon, Lupus, and Ophiuchus are intrinsically 2--4 orders of magnitude
more X-ray luminous than the Sun; however, as a result of the large
($\sim150$ pc) distances to these ``local'' regions of active star formation
and, in many cases, the large X-ray absorbing column depths characteristic
of the dark clouds in which the PMS stars reside, these stars' X-ray fluxes
rarely exceed $10^{-12}$ ergs/cm$^2$/s except during strong flares.  Thus,
it is difficult to perform X-ray spectroscopy of individual pre-main
sequence stars that are located in well-studied regions of star formation.

As a consequence of these limitations, the X-ray emission mechanisms of PMS
stars remain poorly understood. However, the available evidence suggests
that coronal activity is the source of X-rays from low-mass PMS stars.
Until quite recently, most of the evidence was indirect and based on, e.g.,
the observed continuum of X-ray activities from pre-main sequence through
early main sequence evolution and the generally large rotation rates and
X-ray-flaring behavior of T Tauri stars (Kastner et al.\ 1997 [hereafter
KZWF]; Skinner et al.\ 1997; and references therein). These observations
support a scaled-up solar dynamo model underlying the X-ray emission from
low-mass PMS stars (e.g., Carkner et al.\ 1996). With the advent of the {\it
ASCA} X-ray satellite observatory, direct spectroscopic evidence also is
accumulating in support of a coronal origin for the X-rays (e.g., Carkner et
al.; Skinner et al.; Skinner \& Walter 1998).
Generally, {\it ASCA} spectra of PMS stars obtained to date can be
well modeled as arising in coronal plasma with a bimodal temperature
distribution characterized by temperatures
of $\sim3-30$ MK. Thus, \ASCA\ observations have provided new impetus to the
study of PMS stellar X-ray sources.

To better understand what role X-ray emission may play in the chemical and
physical evolution of circumstellar material, and thereby test the
predictions of recent theory (Glassgold, Najita, \& Igea 1997; Shu et al.\
1997), we must obtain high-quality X-ray spectra of individual PMS stars
that are directly analogous to the primordial Sun and solar system. Such
data can also provide probes of intervening circumstellar material, via
the spectral signature of absorption of soft X-rays. With a few exceptions
(e.g., Skinner \& Walter 1998), however, \ASCA\ investigations have focused
either on deeply embedded protostars whose nature and evolutionary status is
uncertain, or on highly active, weak-lined TTS. Classical TTS --- i.e., TTS
surrounded by massive disks that are the likely sites of planet formation
--- have largely been neglected by \ASCA\ studies of PMS stars, mostly due
to the difficulty in identifying suitably X-ray bright candidates.

The classical TTS TW Hya, noteworthy for its large projected distance from
the nearest dark cloud (Rucinski \& Krautter 1983), is a particularly
interesting object in this regard. KZWF established that TW Hya belongs to a
small group of T Tauri star (TTS) systems likely comprising a physical
association (the TW Hya Association; hereafter TWA). These TTS systems form
an exceptionally bright group in X-rays compared with TTS in Taurus and
Chamaeleon, suggesting they are quite close to the Earth. Indeed, on the
basis of \ROSAT\ X-ray and other data, KZWF concluded that the TWA stars are
10--30 Myr old and thus, based on theoretical PMS evolutionary tracks, the
Association lies a mere $\sim50$ pc distant. Hipparcos distances to
Association members TW Hya and HD 98800 --- 57 pc (Wichmann et al.\ 1998)
and 48 pc (Soderblom et al.\ 1998), respectively --- confirm these results,
although there is growing evidence that the ages of most of the Association
stars lie in the range 5 to 15 Myr (Webb et al.\ 1998; Weintraub et
al. 1999). Hence the TW Hya Association, consisting of a small population of
T Tauri stars with no known associated cloud, probably represents the
nearest region of recent star formation. Recently, Webb et al.\ 
identified another five PMS/TTS systems that are likely members of the TWA.

Though all of the late-type TWA members
display various TTS characteristics, such as H$\alpha$ emission, Li
absorption, and/or IR excesses, TW Hya is is the only TWA star to display
both strong H$\alpha$ (Rucinski \& Krautter 1983) and submillimeter
continuum emission (Weintraub, Sandell, \& Duncan 1989). It is also 
the only known CO emission line source among the TWA stars (Zuckerman
et al.\ 1995) and, furthermore, displays submillimeter emission lines of
$^{13}$CO, HCN, CN, and HCO$^+$ (KZWF), despite the apparent lack of 
interstellar molecular gas in its vicinity. KZWF conclude that, given the
narrow ($\sim0.7$ km s$^{-1}$) observed molecular emission line widths, TW
Hya possesses a dusty molecular disk that is viewed nearly
face-on. Considering the $\sim20$ Myr age of TW Hya, this disk may closely
resemble the solar nebula at a time during, or shortly after, the formation
of Jupiter.

Among the small but growing number of PMS solar analogs known to possess
circumstellar, molecular disks, TW Hya is the brightest in X-rays.
Furthermore, it is found in relative isolation, with no nearby X-ray sources
or intervening dark cloud material, and we likely view its star-disk system
nearly pole-on (KZWF). For these reasons, TW Hya presents an excellent
subject for X-ray spectroscopy with \ASCA, which features X-ray spectral
resolution and spectral coverage superior to that of {\it ROSAT} but spatial
resolution inferior to that of its immediate predecessor. In this paper, we
present and analyze \ASCA\ data we acquired for TW Hya and also
re-analyze archival \ROSAT\ pointed observations (Hoff et al.\ 1998). The
results of this analysis provide constraints for models of X-ray irradiation
of circumstellar molecular gas and, more generally, serve as a starting
point for inferences concerning the primordial solar X-ray spectrum at the
epoch of Jovian planet formation. The results also shed light on the
well-documented but puzzling optical variability of TW Hya.

\section{Observations}

\subsection{ASCA}

TW Hya was observed with the two Solid-state Imaging Spectrometers (SIS) and
two Gas Imaging Spectrometers (GIS) aboard \ASCA\ on 1997 June 26-27. Both
types of detectors are sensitive from $\sim$0.5 keV to $\sim$10 keV. The two
SIS instruments afford superior broad-band sensitivity and spectral
resolution ($E/\Delta E \sim 30$ at 6.7 keV), while the two GIS are somewhat
more sensitive than the SIS at higher energies. For these observations, 2 of
the 4 available CCDs in each SIS detector were active.  This configuration
allowed us sufficient off-source detector field of view to determine
background count rates and spectra, while limiting the reporting of spurious
events from off-source CCDs.

Data reduction was accomplished via standard processing software provided by
the High Energy Astrophysics Science Archive Research Center (HEASARC).  We
used standard screening criteria (via the HEASARC XSELECT software) to screen
event data on the basis of Earth elevation angles, particle background
levels, and event CCD pixel distributions, and to reject hot and flickering
CCD pixels, thereby producing lists of valid X-ray events.  To maximize
signal to noise ratio, we have combined data telemetered at ``medium'' and
``low'' bit rates with data telemetered at ``high'' bit rate. The resulting
SIS dataset may, in principle, exhibit slightly degraded spectral resolution
relative to a dataset constructed from the ``high'' bit rate data alone,
since data obtained in ``medium'' and ``low'' bit rates cannot be corrected
for the so-called ``echo effect'' in the SIS CCDs. We cannot discern any
such degradation in the combined dataset relative to that of the ``high''
bit rate data alone, however, and hence the inclusion of ``medium'' and
``low'' bit rate data improves the analysis presented below.  Resulting net
(post-screening) exposure times were 31.1 ks for each SIS detector and 31.4
ks for each GIS detector, out of the total observation duration of $\sim94$
ks ($\sim1.1$ days).

Within the $\sim40'$ diameter GIS and $\sim11'\times22'$ SIS fields, TW Hya
is the only X-ray source detected. We extracted events in $4'$ (SIS) and
$6'$ (GIS) radii circles centered on the position of the source, to
determine count rates and to accumulate spectra (\S
\ref{sec:ASCAspectra}). We also extracted events in off-source regions of
each detector, again using extraction radii of $4'$ (SIS) and $6'$ (GIS), to
obtain representative background count rates and spectra. These background
regions were located $\sim12'$ away from the source regions and hence are
very unlikely to be contaminated with source photons, given the modest count
rates of TW Hya. Results for count rates are listed in Table 1. 

\subsection{ROSAT}

TW Hya was the target of a pointed \ROSAT\ Position Sensitive Proportional
Counter (PSPC) observation on 1991 December 12, and is also included in the
\ROSAT\ All-Sky Survey (RASS) Bright Source Catalog (Voges et al.\ 1998). Here
we re-analyze the pointed PSPC data, which were summarized in KZWF and
subsequently presented in more detail by Hoff et al.\ (1998).  The X-ray
sensitivity of the PSPC plus \ROSAT\ X-ray telescope extends from about 0.1
keV to 2.4 keV, with energy resolution of $\sim40$\% at $\sim1$ keV. We find
that a total of 2100 counts were obtained for TW Hya in 6.8 ks of
detector live time. This count rate, 309$\pm$7 counts ks$^{-1}$,
is consistent with the count rate listed in Table 2 of Hoff et
al. The background count rate, as estimated from off-source
regions of the PSPC image, was $<6$ counts ks$^{-1}$.

\section{Results}

\subsection{ROSAT}
\label{sec:ROSATspec}

%\subsubsection{PSPC spectrum}

\begin{figure}
\label{fig:PSPC}
\plotone{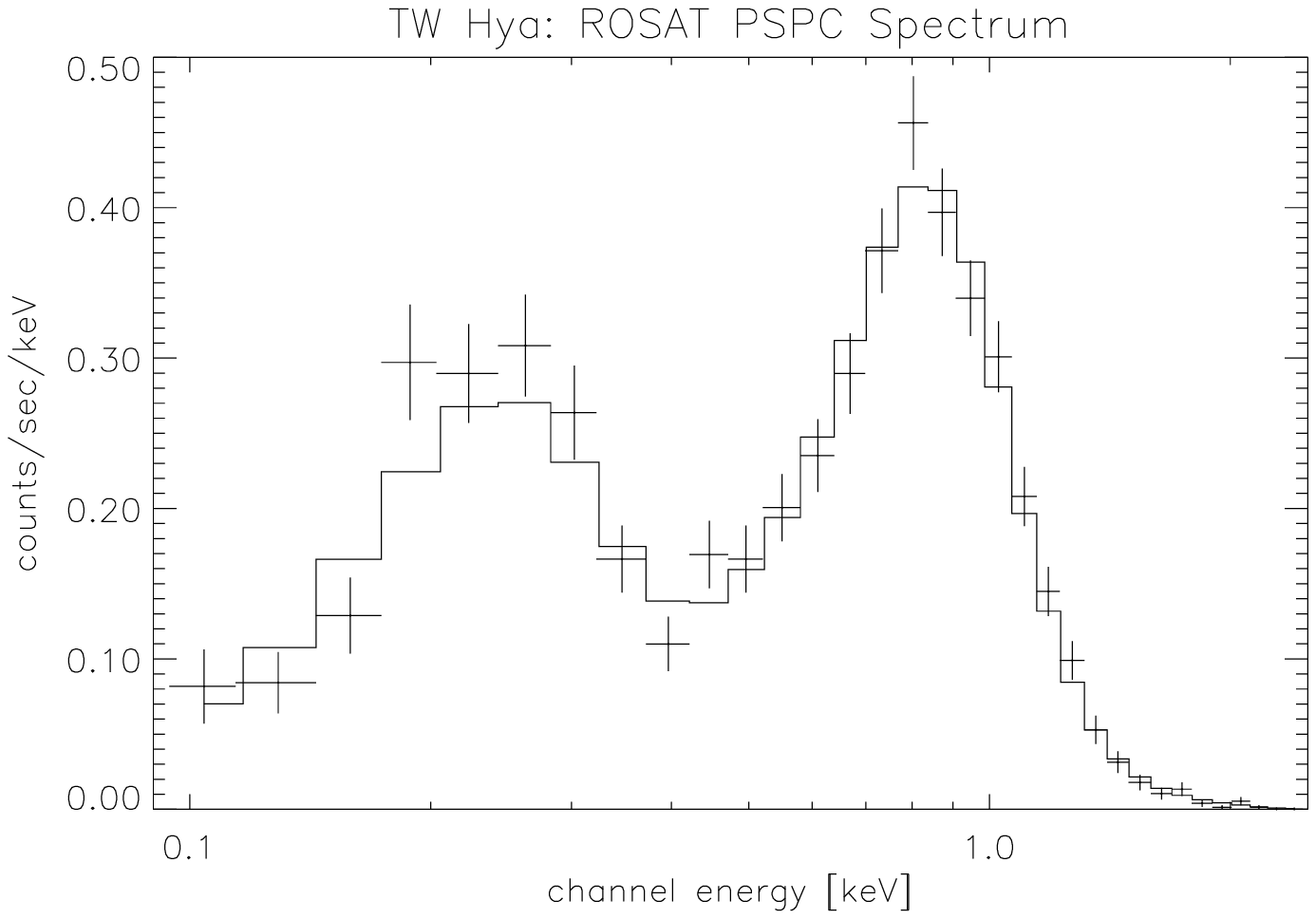}
\caption{
\ROSAT\ PSPC spectrum measured toward TW Hya (crosses), with the best-fit
Raymond-Smith plasma model overlaid (histogram). The depression in both data
and model near 0.4 keV is a result of absorption by the \ROSAT\ mirror.
}
\end{figure}

The PSPC spectrum, binned into 34 energy channels, is displayed in Fig.\
\ref{fig:PSPC} (see also Fig.\ 2 of Hoff et al.\ 1998).  The spectrum is
double-peaked, with peaks near 0.25 and 0.9 keV, and is similar to PSPC
spectra of certain Taurus-Auriga T Tauri stars that are located along lines
of sight with relatively little intervening cloud material (Carkner et al.\
1996). Our tests of fits of pure continuum emission models, such as
blackbody or thermal bremsstrahlung emission, indicate that such models
cannot match the observed shape of the PSPC spectrum peak near 0.9 keV. Pure
continuum models are also ruled out, somewhat more emphatically, by the
\ASCA\ data (\S \ref{sec:ASCAspectra}). Thus, to fit the PSPC spectrum, we
adopt the popular model for X-ray emission from active stars, i.e.,
Raymond-Smith (R-S) coronal plasma emission suffering intervening absorption
due to circumstellar and/or interstellar gas.
%(e.g. Casanova et al.\ 1995; Carkner et al.). 
The R-S models assume collisional ionization equilibrium, and include both
emission lines and continuum contributions for a given temperature and set
of elemental abundances (Raymond \& Smith 1976). We attempted fits of both
single-component and two-component R-S models, with solar abundances. After
convolving with the ROSAT/PSPC spectral response, we find that the latter
model, consisting of a linear combination of R-S models at two different
temperatures, provides a superior fit to the PSPC data. Results for the
best-fit R-S model parameters are listed in Table 2, and the model (folded
through the PSPC spectral response) is overlaid on the data in Fig.\ 1. The
best-fit characteristic temperatures of $kT_1 = 0.14$ keV ($T_1 = 1.7$
MK) and $kT_2 = 0.85$ keV ($T_2 = 9.7$ MK) for the soft and hard
emission components, respectively, are similar to results obtained by Hoff
et al.\ (1998). 

%\subsubsection{PSPC light curve; variability}

A light curve extracted from the pointed PSPC data (not shown)
indicates that, in contrast to the \ASCA\ observations (\S \ref{sec:ASCAlc}),
the X-ray flux from TW Hya was nearly constant during the \ROSAT\ observing
intervals. However, the RASS count rate (570$\pm$40 counts ks$^{-1}$; KZWF)
is approximately double that of the pointed observations. Our event
extraction radius for the pointed data is sufficiently large to rule out
differences in extraction radii as the origin of the different RASS and
pointed count rates. Thus it appears that, like typical TTS, TW Hya probably
is variable in X-rays. 

\subsection{ASCA}

\subsubsection{Spectral analysis}
\label{sec:ASCAspectra}

\begin{figure}
\label{fig:SISRS}
\plotone{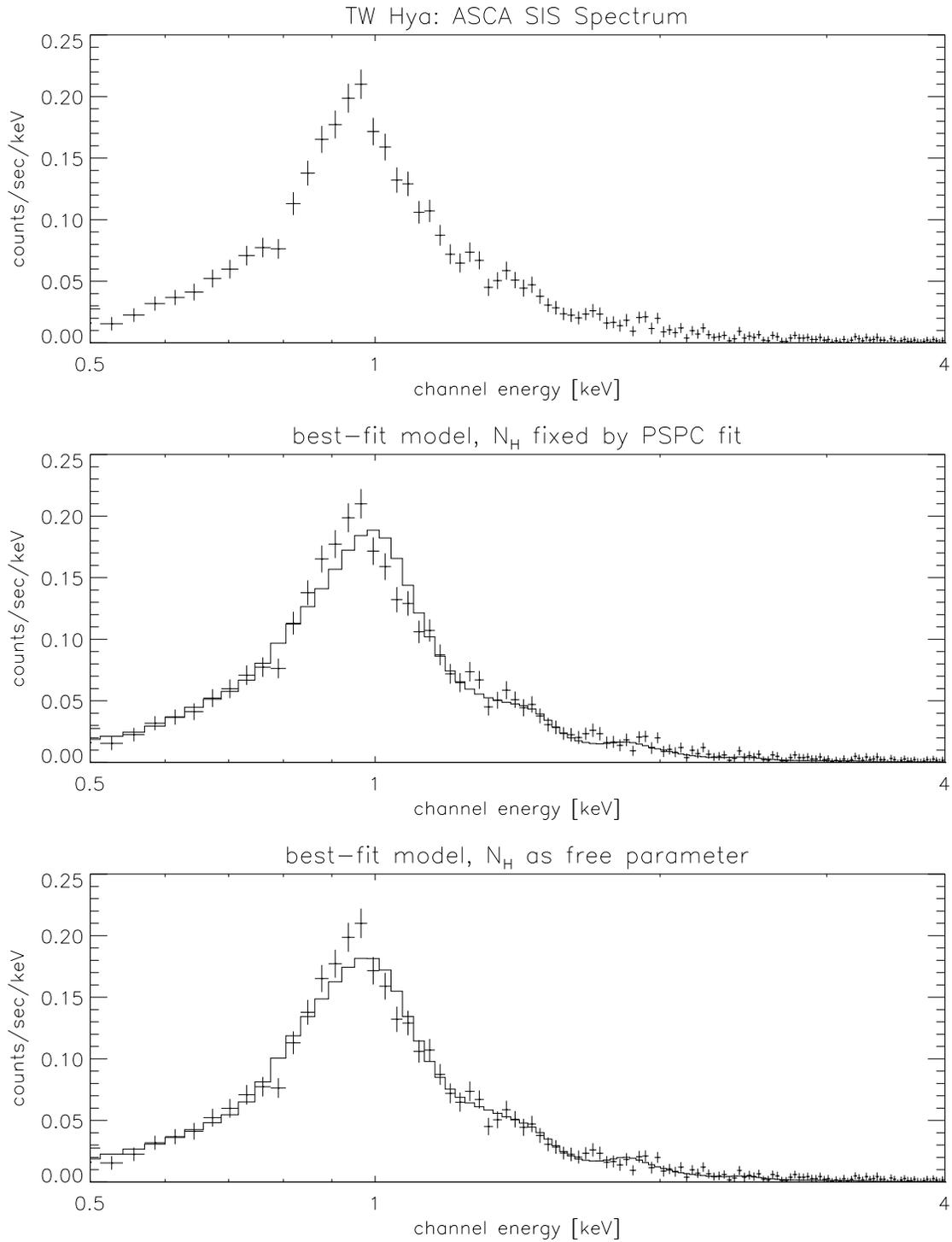}
\caption{
Background-subtracted \ASCA\ SIS0 + SIS1 spectrum (top panel). The lower two
panels show these data (crosses) overlaid with best-fit two-component
Raymond-Smith thermal plasma models (histograms), with absorbing column
fixed by the results of PSPC fitting (middle panel) and with absorbing
column as a free parameter (bottom panel).
}
\end{figure}

We extracted source spectra from the screened SIS0 and SIS1 event lists
using uniform bins of width 29 eV for both
instruments. Although the count rates of the two instruments differ by
$\sim30$\% due to different detector sensitivities (Table 1), the resulting
SIS0 and SIS1 spectra appear very similar. In particular, source spectra
obtained with both instruments peak at about 0.95 keV and display plateaus
or ``shoulders'' of emission on either side of this peak; these shoulders
are centered at about 0.75 keV and 1.3 keV. Other features are also well
reproduced in the corresponding source and background spectra, and tests of
independent fits to the SIS0 and SIS1 source spectra yield similar results
for the two datasets for model parameters such as emitting region
temperature and intervening absorbing column (see below), suggesting the
SIS0 and SIS1 data can be combined without compromising spectral
response. Hence, to maximize signal-to-noise, we averaged the SIS0 and SIS1
source spectra, and then subtracted the average of the SIS0 and SIS1
background spectra from the resulting source spectrum. A weak high-energy
tail that is present in the ``raw'' source spectra is also present in the
background spectra, suggesting it is not intrinsic to the source; this
feature largely disappears from the combined SIS spectrum of TW Hya once
background is subtracted (Fig.\ \ref{fig:SISRS}). We performed all
subsequent spectral analysis on this combined, background-subtracted SIS0
and SIS1 spectrum (which we hereafter refer to as the TW Hya SIS spectrum),
using a hybrid spectral response matrix constructed from a weighted average
of the SIS0 and SIS1 response matrices for the appropriate regions of the
detectors.

Attempts to fit pure continuum models, such as blackbody emission and
thermal bremsstrahlung emission, fail to reproduce the relatively sharp peak
in the spectrum near 0.95 keV.
Significantly improved fits
are obtained once line emission is included. In particular, R-S thermal
plasma models with characteristic temperatures $\sim 10$ MK (and solar
metallicities) reproduce the intensity and overall shape of the
observed 0.95 keV peak (Fig.\ \ref{fig:SISRS}). We therefore conclude that this
feature is dominated by a complex of lines generated by highly ionized species
of Fe, as these lines dominate the spectra of $\sim10$ MK R-S models in the
spectral region near $\sim1$ keV.

We find that single-component R-S models are not able to simultaneously fit
the 0.95 keV peak and the low-energy ($<0.7$ keV) tail in the SIS spectrum.
This residual tail of emission suggests the presence of an additional,
softer component. As the presence of a soft component is also required to
fit the PSPC spectrum of TW Hya (\S \ref{sec:ROSATspec}; Hoff et al.\ 1998),
we attempted fits with two-component R-S models. We did not attempt to use
the PSPC spectral fitting results to constrain the temperatures of the R-S
model fits to the \ASCA\ data, in light of the $\sim5.5$ year time interval
between the \ROSAT\ and \ASCA\ observations. Over this interval there is the
possibility of variations in the physical conditions of the X-ray emitting
region(s) and/or in the distribution of absorbing material along the line of
sight to the star. If we use the results of PSPC spectral analysis to
constrain the absorbing column toward TW Hya at $N_H = 5\times10^{20}$
cm$^{-2}$, we find characteristic temperatures of $\sim2.4$ MK and $\sim12$
MK, respectively, for the soft and hard components in the \ASCA\ spectrum
(Table 2; Fig.\ \ref{fig:SISRS}). If, on the other hand, we allow $N_H$ to
be a free parameter, we find from the SIS spectrum characteristic
temperatures that are very similar to those obtained from our independent
fit to the PSPC data, albeit with substantially larger emission measures for
both soft and hard components (since the best-fit $N_H$ is larger than that
obtained from the fit to the PSPC spectrum). The fit with $N_H$
unconstrained also represents a marginal improvement over that obtained by
holding $N_H$ fixed at $5\times10^{20}$ cm$^{-2}$, particularly in the
$\sim1$ keV spectral region. The results of the ASCA/SIS fit with $N_H$
unconstrained would suggest that $N_H$ was larger by a factor of $\sim6$ at
the epoch of the \ASCA\ observations.

%\subsubsection{Spectra and spectral analysis: GIS}

\begin{figure}
\label{fig:GISRS}
\plotone{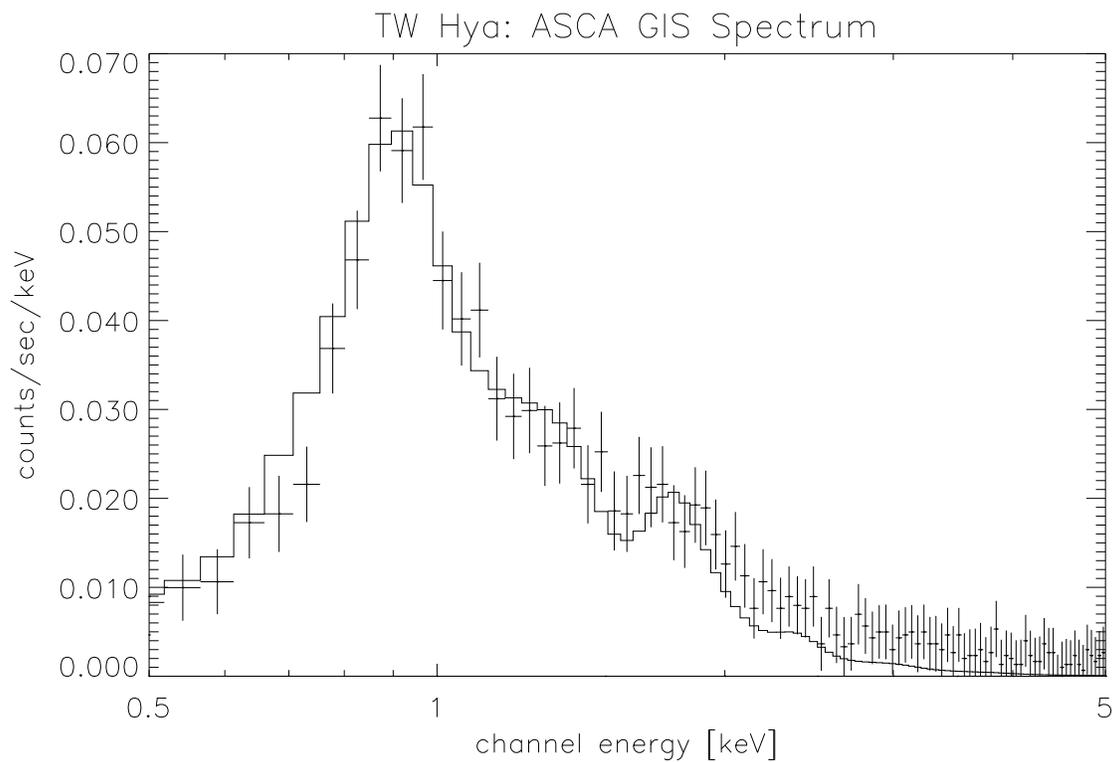}
\caption{
Background-subtracted \ASCA\ GIS2 + GIS3 spectrum (crosses), with
Raymond-Smith thermal plasma model (histogram) overlaid. The characteristic
temperatures and $N_H$ of this model are identical to the model overlaying
the SIS data in the bottom panel of Fig. 2.
}
\end{figure}

We followed a similar procedure to perform spectral analysis of GIS data.
Source spectra were extracted from the screened GIS2 and GIS3 event lists
using uniform bins of width 47 eV. These spectra were combined, and the
combined background spectrum was subtracted from this combined source
spectrum to produce the final GIS spectrum (Figure 3). This spectrum, like
the SIS spectrum, displays a sharp peak at $\sim1$ keV. A hybrid instrument
response matrix was constructed for purposes of spectral fitting. Fits to
the GIS spectrum were constrained by the results of PSPC and SIS model
fitting. Specifically, we used these results to obtain two separate fits to
the GIS spectrum, with $N_H$ fixed at $5\times10^{20}$ cm$^{-2}$ and
emission temperature fixed at $kT = 0.98$ keV, and with $N_H$ fixed at
$2.9\times10^{21}$ cm$^{-2}$ and emission temperature fixed at $kT = 0.86$
keV. We did not include a soft component corresponding to those used in the
PSPC and SIS model fitting, in light of the poor soft X-ray response of the
GIS detectors. With these constraints, we obtain good fits to the GIS
spectrum (Table 2; Fig.\ \ref{fig:GISRS}) at energies $<2$ keV, provided we
apply an offset of 0.1 keV in the energy scale of the GIS response
function. These fits suggest the presence of an additional, harder emission
component in the GIS spectrum (Fig.\ \ref{fig:GISRS}), although due to the
small count rates at channel energies $>2$ keV we are unable to determine a
unique emission temperature for this potential, additional component.

\subsubsection{Temporal Analysis}
\label{sec:ASCAlc}

\begin{figure}
\label{fig:SISGISlc}
\plotone{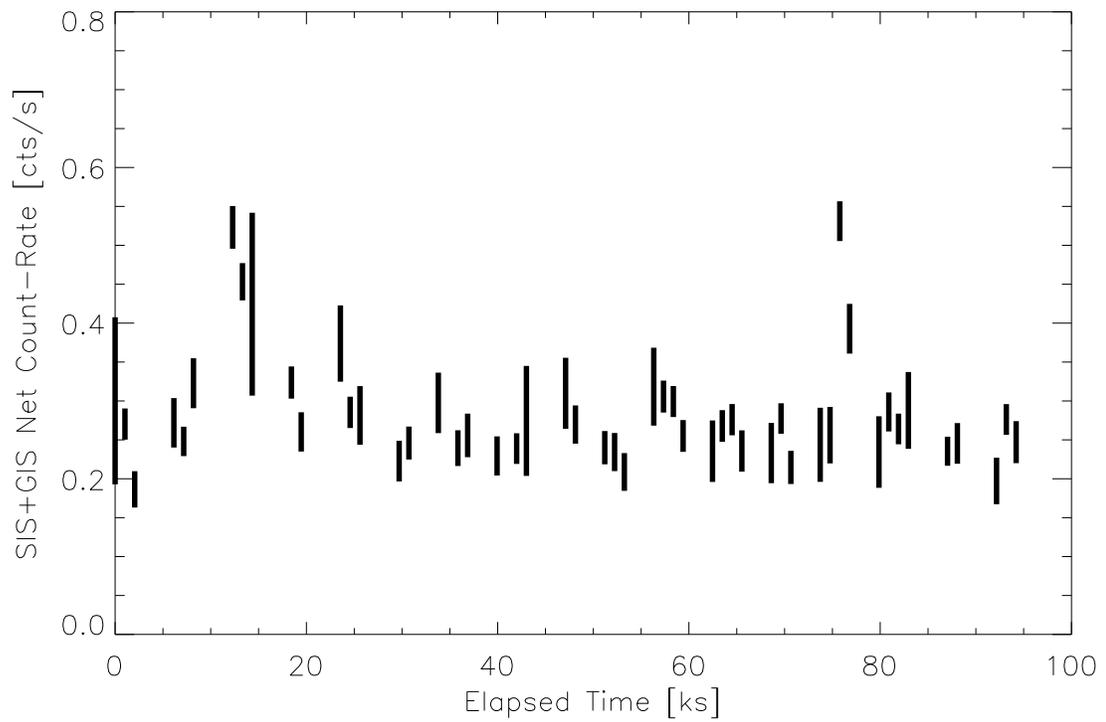}
\caption{
Combined SIS and GIS light curve of TW Hya. The width of the time bins is
1024 s. Symbols show $\pm 1 \sigma$ error bars and are centered on the
measured count rate. Note the short-lived ($\sim2$ ksec duration) ``flares''
at elapsed times of $\sim12$ and $\sim76$ ksec.
}
\end{figure}

A light curve extracted from the SIS and GIS event lists is presented in
Fig.\ \ref{fig:SISGISlc}. The merged GIS and SIS light curve shows some
evidence for sudden, short-lived increases in X-ray flux.  These ``flares''
(detected $\sim0.15$ and $\sim0.87$ days after the start of the observation)
were of duration $\sim1$ hour, separated by about 0.72 days, and represented
a doubling to tripling of the count rate. The increases were observed
independently in all four detectors; therefore, we are able to rule out
background or instrumental effects as the source of these count rate
fluctuations. Spectra extracted from time intervals at high count rate yield
fit results for model parameters that are the same to within the
uncertainties as those obtained for spectra extracted from the entire
observation interval. Hence it appears there were no significant changes in
the X-ray spectrum of TW Hya during the periods of increased X-ray flux,
although the relatively low signal-to-noise ratios of \ASCA\ spectra
obtained during these periods preclude a definitive comparison.

The apparent modest flaring detected in the \ASCA\ observations of TW Hya
stands in contrast to the strong, long-lived flares that have been detected
during \ASCA\ observations of deeply embedded protostars (Koyama et al.\
1996) as well as of magnetically active, weak-lined TTS (Skinner et al.\
1997). These flares are characterized by an order of magnitude or more
increase in X-ray flux accompanied by a significant hardening of the X-ray
spectrum, and have durations of order days. Nevertheless, flaring activity
seems a more plausible explanation for the \ASCA\ count rate increases
observed for TW Hya than does, e.g., rotational modulation of coronal
features (\S 4.3).

\section{Discussion}

\subsection{The Soft, ``Quiescent'' X-ray Spectrum of TW Hya}

Comparison of the results of model fits to the \ROSAT\ and \ASCA\ spectra
(Table 2) indicates that the characteristic emitting region temperatures and
X-ray luminosity of TW Hya changed little between the epochs of
the \ROSAT\ and \ASCA\ observations, although they were obtained about 5.5
years apart. These results indicate that the \ASCA\ and \ROSAT\ spectra are
generally characteristic of a ``quiescent'' state of TW Hya, although we
apparently detected relatively modest, short-lived flares during the ASCA
observation. Evidently, the quiescent X-ray spectrum of TW Hya ($kT \sim
0.85$ keV for the ``hard'' component) is much softer than the quiescent
X-ray spectra of embedded protostars ($kT \sim 6$ keV; Koyama et al.\ 1996)
or of relatively young TTS ($kT \sim 3$ keV for the hard components; Skinner
et al.\ 1997, Skinner \& Walter 1998). Given KZWF's conclusion that TW Hya
is unusually ``old'' for a classical TTS, this comparson suggests a trend
toward softer X-ray emission as a TTS settles onto the main sequence (see
also Skinner \& Walter 1998). This trend is evidently accompanied by only a
modest decrease, if any, in X-ray luminosity; indeed, the ratio of X-ray to
bolometric luminosity increases throughout a late-type star's pre-main
sequence evolution (KZWF). Further X-ray observations may establish whether
or not TW Hya exhibits powerful X-ray flares such as have been observed for
many other TTS, whether its apparent short-term variability is periodic, and
whether its ``quiescent'' X-ray spectral distribution varies with time.

\subsection{X-ray Ionization of the TW Hya Molecular Disk}

The \ASCA\ data likely well characterize the X-ray spectrum that is incident
on surface or subsurface layers of the circumstellar molecular disk
surrounding TW Hya. As described by Glassgold et al.\ (1997),
X-ray radiation incident on a circumstellar molecular disk can partially
ionize the disk if the incident radiation is sufficiently intense and
hard. The position-dependent ionization rate and resulting electron
fractions, though small, may be sufficient to produce stratified disk
instabilities that can govern accretion onto the star as well as
protoplanetary accretion at the disk midplane (Glassgold et al.). Also,
X-ray ionization of molecular gas may play a central role in determining
circumstellar chemistry and, in particular, may explain the enhanced
abundance of HCO$^+$ measured at TW Hya (KZWF).  Thus, the X-ray emission
from TW Hya could be an important factor in determining the chemical
composition of protoplanetary gas and may play a central role in the
formation of any protoplanets themselves, provided there is sufficient X-ray
flux to effectively ionize molecular material in the disk.

The Glassgold et al.\ (1997) and Igea \& Glassgold (1999) models are
formulated for hypothetical star-disk systems. It is worthwhile, therefore,
to consider how closely these models actually resemble the TW Hya system,
which represents in many respects the prototypical example of an
X-ray-bright TTS surrounded by a circumstellar, molecular disk. We have
demonstrated above that TW Hya displays a relatively soft X-ray spectrum; of
the spectra considered in the Glassgold et al.\ (1997) models, for example,
the TW Hya spectrum evidently more closely resembles their assumed ``soft''
(1 keV) spectrum than their ``hard'' (5 keV) spectrum. As Glassgold et al.\
point out, the greater penetration of hard X-rays enhances the ionization
rates of the hard model relative to the soft model, for a given X-ray
luminosity $L_x$.  Hence, the ionization rates produced by the X-ray field
of TW Hya fall short of those that would be produced by, e.g., a younger T
Tauri star or protostar of comparable $L_x$ (but harder X-ray spectrum) that
is surrounded by a molecular disk of comparable mass.  However, the X-ray
luminosity of TW Hya is an order of magnitude larger than that assumed for
the central star in the Glassgold et al.\ models.  Therefore, as the X-ray
ionization rate at a given point in the disk is proportional to $L_x$, the
ionization structure of the TW Hya disk may more closely resemble that of
the Glassgold et al.\ 5 keV model than their 1 keV model (see Fig.\ 2 of
Glassgold et al.).  Moreover, the \ASCA\ data demonstrate that, despite its
relatively soft spectrum, $\sim10$\% of the X-ray luminosity of TW Hya, or
$2\times10^{-29}$ ergs s$^{-1}$, emerges at energies $> 2$ keV. Hence
there is clearly ample hard X-ray flux to produce significant X-ray
ionization of the TW Hya molecular disk, according to the Glassgold et al.\
and Igea \& Glassgold models.

The foregoing comparison suggests that a more detailed analysis, along the
lines of that carried out by Igea \& Glassgold (1999) but tailored to
conditions appropriate for TW Hya, would be of great interest. Such an
analysis should take into account the fact that the X-ray spectrum of TW Hya
evidently is dominated by emission lines at energies near 1 keV (\S
\ref{sec:ASCAspectra}), and that its molecular disk mass ($\sim
7\times10^{28}$ g, or $\sim12$ Earth masses; KZWF) is a factor $\sim100$
smaller than that of the ``canonical'' minimum mass solar nebula. The latter
constraint implies large ionization rates near the TW Hya disk mid-plane,
relative to the ionization rates characteristic of models tailored to the
minimum mass solar nebula.

\subsection{$N_H$: Constraints on Circumstellar and Viewing Geometries}

The range in absorbing column indicated by the fits to the PSPC and SIS
spectra constrains both the distribution of circumstellar material and our
view of the star through this material. Assuming that $N_H$ is dominated by
circumstellar (as opposed to interstellar) extinction at the large neutral H
densities implied by the detection of HCN line emission ($n_{H_2} > 10^{7}$
cm$^{-3}$; KZWF), the value of $N_H$ derived from the SIS data ($N_H \sim
2.9\times10^{21}$ cm$^{-2}$) suggests the column length along the line of
sight to the star is $l < 3 \times10^{14}$ cm ($l <20$ AU). The PSPC result
for $N_H$ would impose a more stringent upper limit of $l < 5 \times10^{13}$
cm ($l <3$ AU). Based on CO line strengths and ratios, KZWF concluded that
the projected radius of the molecular line emitting region is $\sim60$
AU. Thus, if the material responsible for the absorption measured in the
X-ray spectra of TW Hya resides in the same circumstellar structure as is
responsible for the molecular line emission, then this structure is
flattened and is viewed from high latitude (if viewed edge-on, the TW Hya
disk would have a molecular column density $N_{H_2} > 10^{22}$
cm$^{-2}$). On the other hand, the presence of a considerable column depth
of ionized gas is reflected in the fact that TW Hya displays a huge
H$\alpha$ equivalent width (e.g., KZWF). Such a large H$\alpha$ EW likely
results from a large mass loss rate (e.g., Bouvier et al.\ 1995) and this
material most likely is ejected along the poles of the star. Hence the X-ray
absorption may be produced in an extended, ionized stellar wind. In either
case, the results for $N_H$ appear to support the conclusion that the dense,
relatively cold ($T\sim100$ K) region of circumstellar molecular gas and
dust detected in the millimeter and submillimeter wavelength regime resides
in a flattened structure surrounding TW Hya that is viewed nearly pole-on
(Zuckerman et al.\ 1995; KZWF).

\subsection{Variability of Optical Photometry and X-rays}

Photometry published in Rucinski \& Krautter (1983) indicated that TW Hya
displays large-amplitude ($\sim0.5$ mag) variability over short (1-day)
timescales. Herbst \& Koret (1988) derived periods of 1.28 days and 1.83
days from their own $UBV$ photometric monitoring and from the older dataset
of Rucinski \& Krautter, respectively. Herbst \& Koret attributed the
variability to rotational modulation of a hot spot. More recently, Mekkaden
(1998) derived a period of 2.196 days from independent photometric
monitoring and claimed that this longer period satisfies all available
photometry, including the monitoring data analyzed by Herbst \&
Koret. However, the phases of maximum and minimum light can be seen to shift
randomly in the folded light curves for various observing epochs presented
by Mekkaden.

\begin{figure}
\label{fig:Hipp}
\plotone{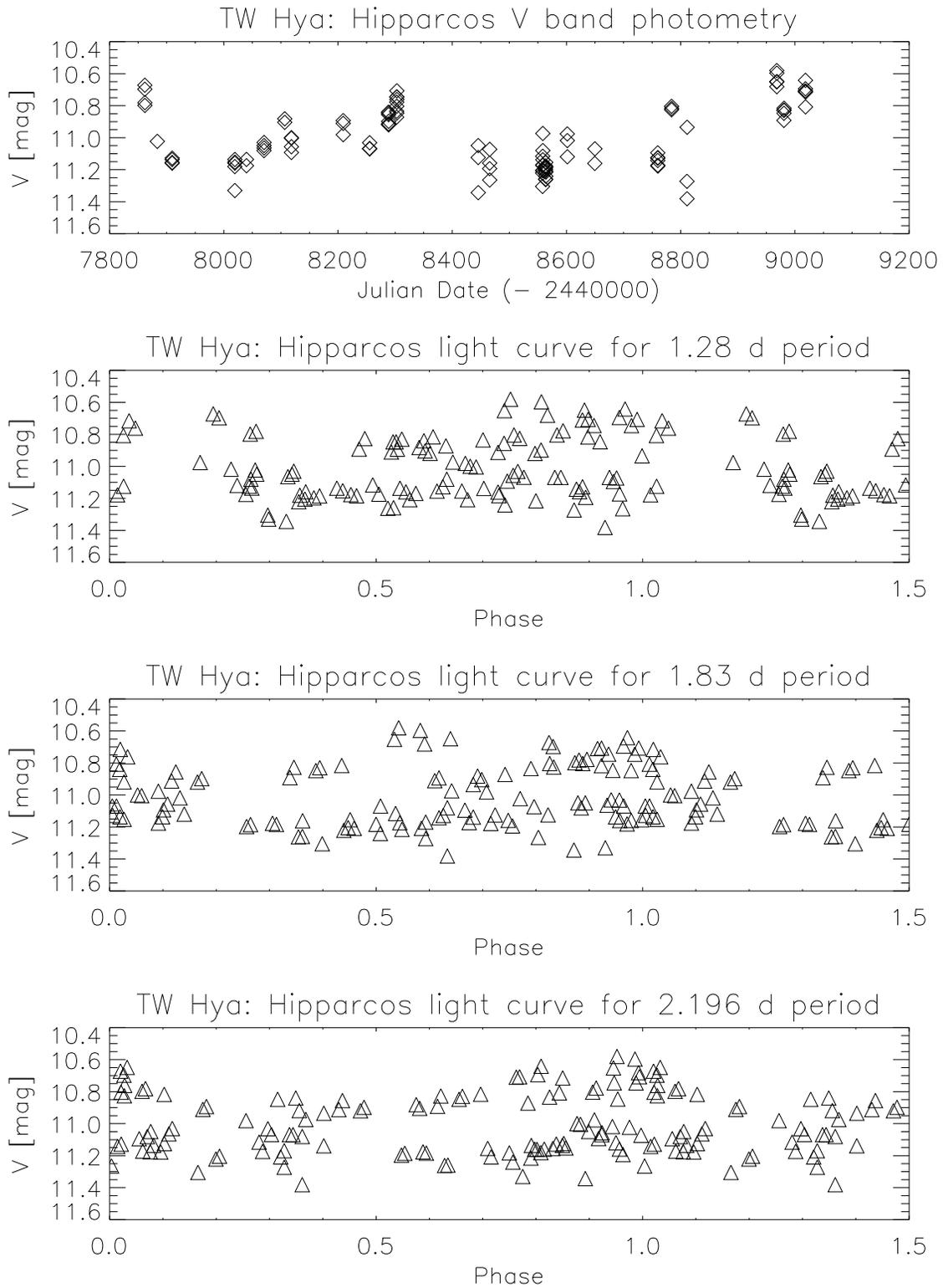}
\caption{
Hipparcos V band light curve of TW Hya (top). The remaining 3 panels show
phase diagrams constructed by folding the Hipparcos V band photometry
according to various periods derived in the literature (see \S 4.4).
}
\end{figure}

Given the apparent inconsistency in the various published results for the
period of TW Hya, we examined archival optical ($V$ band) 
photometry of TW Hya obtained during the course of the Hipparcos astrometry
mission. This photometry is rather sparsely sampled, but extends over a
long temporal baseline encompassing the time period of the various
photometric datasets presented by Mekkaden (1998). In addition to revealing
$V$ fluctuations of $\sim0.2$ mag over the course of a day or so, the
Hipparcos data well illustrate the larger, longer-term optical variability
of TW Hya (Fig.\ \ref{fig:Hipp}a). Indeed, these data include the brightest
and faintest $V$ magnitudes (minimum $V=10.58\pm0.03$; maximum
$V=11.38\pm0.04$), and hence display the largest amplitude of $V$ band
variation ($\Delta V = 0.8$ mag), observed thus far for TW Hya.

We have folded the Hipparcos V photometry according to the various periods
cited in the literature (Fig.\ \ref{fig:Hipp}b-d). There is considerable
scatter in each of these folded lightcurves, and no regular variation can be
discerned. We conclude that none of the published periods well describes the
Hipparcos photometry. The Hipparcos data therefore cast doubt on the
existence of well-defined, short-period optical variability of TW Hya, such
as would be due to rotational modulation.

A general problem with the interpretation of the optical variability in
terms of rotational modulation is the likely viewing geometry. Both Herbst
\& Koret (1988) and Mekkaden (1998) contend that the wavelength dependence
of the variability amplitude of TW Hya is best interpreted as rotation of a
hot (as opposed to cool) spot and, hence, may indicate the presence of an
accretion hotspot at the boundary layer between star and disk. There now
exists independent evidence for the presence of a circumstellar disk
(Weintraub et al.\ 1989; KZWF). However, there is accumulating evidence that
the disk is viewed nearly pole-on (\S 4.3; KZWF), and the measured $v
\sin{i} < 15$ km s$^{-1}$ (Franchini et al.\ 1992; Webb 1998, private comm.)
suggests the star itself is also viewed from a high latitude. Even if the
system is viewed nearly, but not precisely, pole-on, a model that interprets
the variation of $UBV$ radiation in terms of the rotation of an accretion
hotspot requires either that the hotspot is very near the star's equator or
that the star's rotation axis is not well aligned with the disk
axis. Neither idea is favored by contemporary protostellar accretion theory
(e.g., Shu et al.\ 1997).

Accepting for the moment that Herbst \& Koret (1988) and Mekkaden (1998)
have detected rotational modulation in the light curve of TW Hya, any of the
periods derived by these investigations would make TW Hya one of the fastest
rotating classical TTS known. Generally, however, its somewhat erratic
photometric behavior resembles that of other classical TTS stars, many of
which exhibit temporal variations in and disappearances/reappearances of
photometric periodicity (Bouvier et al.\ 1995). Like Herbst \& Koret and
Mekkaden, Bouvier et al.\ interpret the variability of the stars in
their classical TTS sample in terms of rotational modulation of hot spots.
However, the changes, disappearances, and reappearances of periodicity in
the photometry of TW Hya and other classical TTS suggest that quasi-random
flaring, perhaps caused by short-lived, small-scale accretion events, might
better explain their short timescale optical variability. Similar processes
may be responsible for the X-ray outbursts seen by \ASCA\ (Fig.\
\ref{fig:SISGISlc}).

The longer-term optical variability of TW Hya, on the other hand, may be
related to changes in X-ray absorbing column (\S 3.2.1).  The fits to the
PSPC and SIS spectra indicate that $N_H$, and hence $A_V$, may have changed
by a factor $\sim6$ over the intervening $\sim5.5$ yr. Based on conversion
factors from $A_V$ to $N_H$ listed in Ryter (1996) and Neuhauser, Sterzik,
\& Schmitt (1995), the results for $N_H$ correspond to $A_V \sim 0.25$ and
$A_V \sim 1.5$ for the epochs of the ROSAT and \ASCA\ data,
respectively. Given the star's UV and near-IR excesses, it is difficult to
determine reddening via e.g.\ observed vs.\ intrinsic $B-V$ color (Rucinski
\& Krautter 1983); hence these results for $A_V$ serve as perhaps the most
reliable indications yet obtained of the magnitude and variability of visual
extinction toward TW Hya.

The large range inferred for $A_V$ raises the possibility that the long
timescale, large amplitude optical variability of TW Hya is due to variable
circumstellar obscuration. In this regard it is intriguing that the
Hipparcos data appear to show long-term ($\sim600$ day) periodicity
characterized by steep declines in $V$ followed by more gradual
recoveries. This variation is reminiscent of, though not nearly as dramatic
as, the behavior of R CrB stars, which vary by many magnitudes in the
optical due to the formation and then gradual dispersal of dust clouds along
the line of sight to the stars (Clayton 1996). Observations that the
amplitude of variation increases steeply with decreasing wavelength and that
the degree of optical polarization changes with time 
(Mekkaden 1998) are consistent with variable obscuration of the star by
dust. Furthermore the difference in inferred $A_V$ between the epochs of the
\ROSAT\ and \ASCA\ observations (1.2 mag) is on the order of the amplitude
of $V$ band variation measured by Hipparcos (0.8 mag).  On this basis, we
would predict that TW Hya was at or near its faintest at the time of the
\ASCA\ observations (June 1997; J.D. 2450626), as the $A_V$ inferred from
the \ASCA\ data is large.  Unfortunately we are not aware of the existence
of optical photometry that might provide a test of this prediction. However,
at the approximate time of the \ROSAT\ pointed observations (Dec.\ 1991;
J.D. 2448602), for which we infer a relatively small $A_V$ and therefore we
anticipate that TW Hya was quite bright optically, Hipparcos measured only
$V\sim11.0$ (Fig.\ \ref{fig:Hipp}a) --- roughly the median of the Hipparcos
$V$ band measurements. Thus, while the large range of $A_V$ inferred from
the X-ray data is suggestive of the presence of variable obscuration, it is
not clear that these values provide an accurate measure of the relative $V$
magnitude of TW Hya.

In contrast to the episodic dust formation and
dispersal characteristic of R CrB stars, we speculate that the appearance
and gradual disappearance of dust clouds along the line of sight to TW Hya
might be produced by episodes in which infalling disk material is partially
accreted and partially redirected into polar jets or outflows. This
explanation seems consistent with the variability of optical continuum and
H$\alpha$ emission from TW Hya (Mekkaden 1998) as well as its X-ray
variability.  Further contemporaneous X-ray spectroscopy and optical
photometric and polarimetric monitoring would test the hypothesis that the
long-term variability of TW Hya and other classical T Tauri stars is due to
variable circumstellar extinction.

\section{Conclusions}

We have obtained and analyzed \ASCA\ data and re-analyzed archival \ROSAT\
data for the nearby, isolated classical T Tauri star TW Hya. We find that
both \ROSAT\ and \ASCA\ spectra are well fit by a two-component
Raymond-Smith thermal plasma model, with characteristic temperatures of $T
\sim1.7$ MK and $T \sim9.7$ MK and X-ray luminosity of $L_x \sim
2\times10^{30}$ ergs s$^{-1}$. Prominent line emission, likely arising from
highly ionized species of Fe, is apparent in the \ASCA\ spectra at energies
$\sim1$ keV, while the GIS spectrum displays a weak hard X-ray excess at
energies $> 2$ keV.  Although modest flaring is apparent in the \ASCA\ light
curve, the overall similarity of the X-ray spectra inferred from the \ASCA\
and \ROSAT\ observations suggests that the X-ray data obtained to date are
characteristic of a ``quiescent'' TW Hya.

There is evidence that the absorption of the X-ray spectrum of TW Hya may
have changed in the $\sim5.5$ year interval between \ROSAT\ and \ASCA\
observations. The apparent long-term change in $N_H$ and the appearance of
short-lived flares in the \ASCA\ data appear to be related to the large
amplitude optical variability of TW Hya on both long and short timescales;
neither the X-ray nor optical variability appears to be well explained by
rotational modulation. 

Since TW Hya appears to be a single star of age $\sim10$ Myr that is
surrounded by a circumstellar molecular disk (KZWF), the
\ASCA\ spectra offer an indication of the X-ray spectrum
of the young Sun during the epoch of Jovian planet building. These spectra
indicate that the hard X-ray flux from TW Hya ($\sim2\times10^{29}$ ergs
s$^{-1}$ at energies $> 2$ keV) is sufficient to produce substantial
ionization of its circumstellar molecular disk.

Further X-ray spectroscopic and monitoring observations of TW Hya might find
the star in a strong, longer-lived flaring state, and would better establish
whether or not the physical conditions responsible for its ``quiescent''
spectrum are stable. In this regard, we note that TW Hya will be observed by
both the Chandra X-ray Observatory (CXO) and the X-ray Multimirror Mission
(XMM) during their first years of operation\footnote{As of this writing, CXO
was scheduled for launch in summer 1999, while XMM launch was scheduled for
January 2000.}.  Chandra and XMM observations of TW Hya should far surpass
the \ASCA\ and \ROSAT\ results presented here in terms of spectrum quality,
and thus should provide additional insight into the nature of the X-ray
emission from this seminal young, Sun-like star. In particular,
high-resolution (gratings spectrometer) CXO and XMM spectra of TW Hya will
provide critical tests of our conclusions that the X-ray spectrum of TW Hya
is coronal in origin and that the features seen near 1 keV in \ASCA\ and
\ROSAT\ spectra are formed by a blended complex of highly ionized iron
lines. Low-resolution (CCD) spectra resulting from both missions will
provide additional measures of the degree and variability of intervening
circumstellar absorption.

\acknowledgements{Primary support for this work was provided by NASA \ASCA\
data analysis grant NAG5--4959 to M.I.T. Research by J.H.K., D.P.H., and
N.S.S. at M.I.T. is also supported by the Chandra X-ray Observatory
Science Center as part of Smithsonian Astrophysical
Observatory contract SVI--61010 under NASA Marshall Space Flight Center.
Support for research by D.A.W. and J.H.K. was also provided by NASA Origins
of Solar Systems program grant NAG5--8295 to Vanderbilt University.}

\pagestyle{empty}

\end{document}